\documentclass[11pt]{amsart}
\usepackage[dvips,colorlinks=true,linkcolor=blue,urlcolor=red]{hyperref}
\usepackage[light]{}
\usepackage{pstricks,pst-plot}
\numberwithin{equation}{section}

\newcommand{\R}{\mathbb{R}}

\newcommand{\N}{\mathbb{N}}

\newcommand{\Z}{\mathbb{Z}}

\newcommand{\C}{\mathbb{C}}

\theoremstyle{plain}
\newtheorem{theorem}{Theorem}[section]
\newtheorem{Th}[theorem]{Theorem}
\newtheorem{Pro}[theorem]{Proposition}

\newtheorem{Le}[theorem]{Lemma}
\newtheorem{Remark}[theorem]{Remark}
\newtheorem{Def}[theorem]{Definition}

\renewcommand{\epsilon}{\varepsilon}

\newcommand{\smallpagebreak}{\newline}

\author{M. Marx}
\address[M. Marx]{LAGA, U.M.R. 7539 C.N.R.S, Institut Galil{\'e}e,
  Universit{\'e} de Paris-Nord, 99 Avenue J.-B. Cl{\'e}ment, F-93430
  Villetaneuse, France}
\email{\href{mailto:marx@math.univ-paris13.fr}{marx@math.univ-paris13.fr}}
\author{H. Najar}
\address[Hatem Najar]{Département de Mathématiques I.S.M.A.I. Kairouan, Rue Assed Ibn Elfourat 3100 Kairouan, Tunisia }
\email{\href{mailto:hatem.najar@ipeim.rnu.tn}{hatem.najar@ipeim.rnu.tn}}

\begin{document}
\baselineskip=20pt
\renewcommand{\theequation}{\arabic{section}.\arabic{equation}}
\title{On the singular spectrum for adiabatic quasi-periodic Schr{\"o}dinger Operators}

\maketitle

\begin{abstract}
 In this paper we study spectral properties of a family
 of quasi-periodic Schr\"odinger operators on the real line in the
 adiabatic limit. We assume that the adiabatic iso-energetic curve has a real branch that
 is extended along the momentum direction. In the energy intervals where this happens, we obtain an asymptotic
 formula for the Lyapunov exponent and show that the spectrum is
 purely singular. This result was conjectured and proved in a particular case by Fedotov and Klopp in \cite{FEKL1}.
 \end{abstract}

\section{Introduction}
\label{sec:intro} We consider the following Schr\"odinger equation
\begin{equation}
\label{eq:qper}
 (H_{z,\varepsilon}\psi)(x)=-\frac{d^{2}}{dx^{2}}\psi(x)+[V(x-z)+W(\varepsilon
x)]\psi(x)=E\psi(x),\quad x\in\R,
\end{equation}
where $x\mapsto V(x)$ and $\zeta\mapsto W(\zeta)$ are periodic, and
$\varepsilon$ is chosen so that the potential
$V(\cdot-z)+W(\varepsilon\cdot) $ be quasi-periodic. Note that in
this case, the family of equations (\ref{eq:qper}) is ergodic, see
\cite{PasFig}; in this case so its spectrum does not depend on $z$,
see \cite{AvSi83}. The operator $H_{\varphi,\varepsilon}$ can be
regarded as an adiabatic perturbation of the periodic operator
$H_{0}$:
\begin{equation}
\label{eq:opper}
 H_{0}=-\frac{d^{2}}{dx^{2}}+V(x).
\end{equation}
Equation (\ref{eq:qper}) is one of the main models of solid state
physics. The function $\psi$ is the wave function of an electron in
a crystal with impurities. $V$ represents the potential of the
perfect crystal; as such it is periodic. The potential $W$ is the
perturbation created by impurities. In the semiconductors, this
perturbation is slow-varying with respect to the field of the
crystal, \cite{Zi}. It is natural to consider the semi-classical
limit. \newline Let $\mathcal{E}(\kappa)$ be the dispersion relation
associated to $H_0$. Consider the $\textit{complex}$ and
$\textit{real-isoenergy curves}$ $\Gamma_{\mathbb{C}}$ and
$\Gamma_{\mathbb{R}}$ defined by
\begin{equation}
\Gamma_{\mathbb{C}}(E)=\{(\kappa,\zeta)\in \mathbb{C}^2;\
\mathcal{E}(\kappa)+W(\zeta)=E\};
\end{equation}
\begin{equation}
\Gamma_{\mathbb{R}}(E)=\{(\kappa,\zeta)\in \mathbb{R}^2;\
\mathcal{E}(\kappa)+W(\zeta)=E\}.
\end{equation}
Notice that the iso-energy curves $\Gamma_{\mathbb{C}}(E)$ and
$\Gamma_{\mathbb{R}}(E)$ are $2\pi$-periodic in $\zeta$ and $\kappa$
and $\Gamma_{\mathbb{C}}$ is the Riemann surface uniformizing
$\kappa$.
\newline The real iso-energy curve has a well known role for adiabatic problems
\cite{deux}. The adiabatic limit can be regarded as a semi-classical
limit and the Hamiltonian $\mathcal{E}(\kappa)+W(\zeta)$ can be
interpreted as a "classical" Hamiltonian corresponding to equation
(\ref{eq:qper}). \newline When $W$ has in a period exactly one
maximum and one minimum, that are non-degenerate, it is proved in
\cite{FEKL1} that in the energy intervals where the adiabatic
iso-energetic curves are extended along the momentum direction, the
spectrum is purely singular. This result leads to the following
conjecture: in a given interval, if the iso-energy curve has a real
branch (a connected component) that is an unbounded vertical curve,
then in the adiabatic limit, in this interval, the spectrum is
singular. This paper is devoted to prove this conjecture.\newline
Heuristically when the real iso-energy curve is extended along the
momentum axis, the quantum states should be extended in momentum
and, thus localized in the position space. 
 \subsection{Results
and discussions } \label{sec:res} Now, we state our assumptions and
results. \subsubsection{Assumptions on the potentials} We assume
that
\begin{description}
\item[(H1)] $V$ and $W$ are periodic,
\begin{equation}
V(x+1)=V(x),\qquad W(x+2\pi)=W(x).
\end{equation}
\item[(H2)] $V$ is real-valued and locally square-integrable.
\item[(H3)] $W$ is real analytic in the strip $S_{Y}=\{z\in\C\ ;\
|\textrm{Im}\  z|<Y\}$.
\end{description}
We define:
\begin{equation}\label{eq:minmax}
W_{-}=\inf\limits_{t\in\R}W(t),\quad\quad
W_{+}=\sup\limits_{t\in\R}W(t)
\end{equation}
\subsubsection{Assumptions on the energy region}
\label{sec:enerregion} To describe the energy regions where we study
the spectral properties, we consider the periodic Schr{\"o}dinger
operator $H_{0}$ acting on $L^{2}(\R)$ and defined by
(\ref{eq:opper}). \subsubsection{The periodic operator} The spectrum
of (\ref{eq:opper}) is absolutely continuous and consists of
intervals of the real axis, say $[E_{2n+1},\,E_{2n+2}]$ for $n\in\N$
, such that $ E_{1}<E_{2}\leq E_{3}< E_{4}...E_{2n}\leq E_{2n+1}<
E_{2n+2}...$ and $E_{n}\rightarrow + \infty,n\rightarrow +\infty$.
The points $(E_{j})_{j\in\N}$ are the eigenvalues of the
self-adjoint operator obtained by considering $H_0$ defined by
(\ref{eq:opper}) and acting in $L^2([0,1])$ with periodic boundary
conditions (see \cite{Eas:73, ReSi4}). The intervals
$[E_{2n+1},\,E_{2n+2}]$, $n\in\N$, are the {\it spectral
 bands}, and the intervals $(E_{2n},\,E_{2n+1})$, $n\in\N^*$, the
{\it spectral gaps}. When $E_{2n}<E_{2n+1}$, one says that the
$n$-th gap is {\it open}; when $[E_{2n-1},E_{2n}]$ is separated from
the rest of the spectrum by open gaps, the $n$-th band is said to be
{{\it {isolated}}}. The spectral bands and gaps are represented in
figure \ref{fig:bgqm}.

\psset{unit=1em,linewidth=.1}
\begin{center}
\begin{figure}
\begin{pspicture}(-10,-10)(30,10)

\pscurve{->}(2,8.5)(7,9.5)(12,8.5)
\uput[180](1,7.5){$(\mathcal{E})$}
\uput[180](12,7.5){$(k_{0}(\mathcal{E}))$}

\psline(-6,5)(-4,5) \psline(-2,5)(0,5) \psline(2,5)(3,5)
\psdots[dotstyle=*](-6,5)(-4,5)(-2,5)(0,5)(2,5)
\uput[180](-6,4){$E_{1}$}

\uput[180](-4,4){$E_{2}$}
\uput[180](-2,4){$E_{3}$}\uput[180](0,4){$E_{4}$}\uput[180](2,4){$E_{5}$}
\psline[linestyle=dotted]{>-}(-8,5.4)(-7,5.4)(-6.4,5.4)
\psarc[linestyle=dotted](-6,5.4){0.4}{0}{180}
\psarc[linestyle=dotted](-4,5.4){0.4}{0}{180}
\psarc[linestyle=dotted](-2,5.4){0.4}{0}{180}
\psarc[linestyle=dotted](0,5.4){0.4}{0}{180}
\psarc[linestyle=dotted](2,5.4){0.4}{0}{180}
\psline[linestyle=dotted](-5.6,5.4)(-4.4,5.4)
\psline[linestyle=dotted](-3.6,5.4)(-2.4,5.4)
\psline[linestyle=dotted](-1.6,5.4)(-.4,5.4)
\psline[linestyle=dotted](0.4,5.4)(1.6,5.4)
 \psline[linestyle=dotted](2.4,5.4)(3,5.4)

\uput[180](-7.6,6.5){$\gamma_{1}$}

\psline(-6,-5)(-4,-5) \psline(-2,-5)(0,-5) \psline(2,-5)(3,-5)
\psdots[dotstyle=*](-6,-5)(-4,-5)(-2,-5)(0,-5)(2,-5)
\uput[180](-4,-6.5){$E_{2}$}
\uput[180](-2,-6.5){$E_{3}$}\uput[180](0,-6.5){$E_{4}$}\uput[180](2,-6.5){$E_{5}$}
\psline[linestyle=dashed](3,-5.4)(-6,-5.4)

\psarc[linestyle=dashed](-6,-5){0.4}{90}{270}

\psline[linestyle=dashed]{->}(-6,-4.6)(3,-4.6)
\uput[180](2.5,-3.8){$\gamma_{2}$}

\psline(20,-4)(20,7) \psline(12,0)(29,0) \psline(17,0)(17,4)
\psline(14,0)(14,1.5)\psline(23,0)(23,4)\psline(26,0)(26,1.5)
\uput[180](14,-0.6){$-2\pi$}\uput[180](26,-0.6){$2\pi$}
\uput[180](17,-0.6){$-\pi$}\uput[180](23,-0.6){$\pi$}
\psline[linestyle=dashed](12,0.4)(13.6,0.4)
\psline[linestyle=dashed](13.6,0.4)(13.6,1.5)
\psarc[linestyle=dashed](14,1.5){0.4}{0}{180}
\psline[linestyle=dashed](14.4,1.5)(14.4,0.4)

\psline[linestyle=dashed](14.4,0.4)(16.6,0.4)
\psline[linestyle=dashed](16.6,0.4)(16.6,4)
\psarc[linestyle=dashed](17,4){0.4}{0}{180}
\psline[linestyle=dashed](17.4,0.4)(17.4,4)

 \psline[linestyle=dashed]{->}(17.4,0.4)(20,0.4)
 \psline[linestyle=dashed](20,0.4)(22.6,0.4)
\psline[linestyle=dashed](22.6,0.4)(22.6,4)
\psarc[linestyle=dashed](23,4){0.4}{0}{180}
 \psline[linestyle=dashed](23.4,4)(23.4,0.4)
\psline[linestyle=dashed](23.4,0.4)(25.6,0.4)
\psline[linestyle=dashed](25.6,0.4)(25.6,1.5)
\psarc[linestyle=dashed](26,1.5){0.4}{0}{180}
 \psline[linestyle=dashed](26.4,0.4)(26.4,1.5)
 \psline[linestyle=dashed]{>-}(26.4,0.4)(29,0.4)

\psline[linestyle=dotted](20.4,7)(20.4,5)
\psline[linestyle=dotted]{>-}(20.4,5)(20.4,0.4)
\psline[linestyle=dotted](20.4,0.4)(22.6,0.4)
 \psline[linestyle=dotted](22.6,0.4)(22.6,4)
\psarc[linestyle=dotted](23,4){0.4}{0}{180}
 \psline[linestyle=dotted](23.4,4)(23.4,0.4)
\psline[linestyle=dotted](23.4,0.4)(25.6,0.4)
\psline[linestyle=dotted](25.6,0.4)(25.6,1.5)
\psarc[linestyle=dotted](26,1.5){0.4}{0}{180}
 \psline[linestyle=dotted](26.4,0.4)(26.4,1.5)
 \psline[linestyle=dotted]{->}(26.4,0.4)(29,0.4)

\uput[180](18.2,1){$\gamma_{2}$} \uput[180](21.2,5){$\gamma_{1}$}
\uput[180](28,1){$\gamma_{1},\gamma_{2}$}

\end{pspicture}
\caption{Bands, gaps and the quasi-momentum $k$} \label{fig:bgqm}
\end{figure}
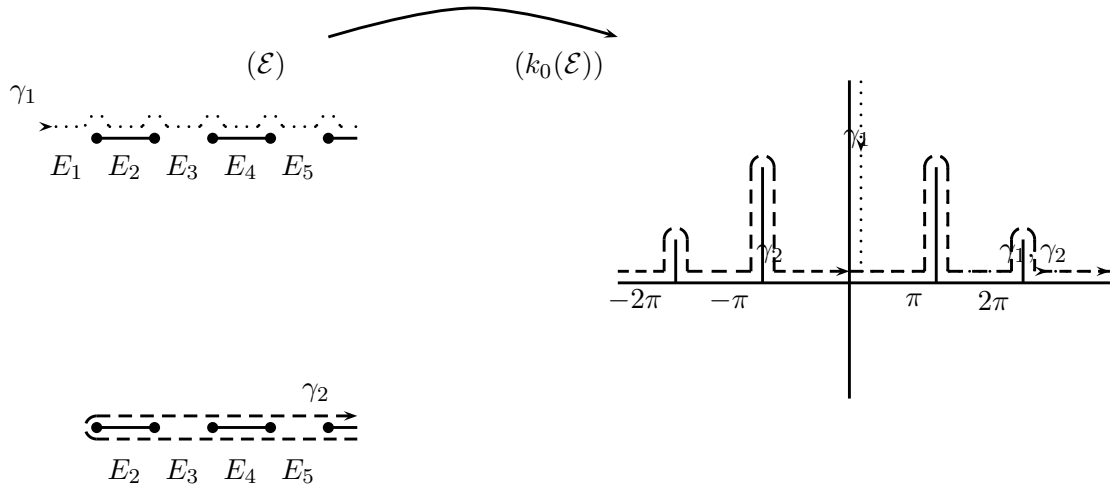
\end{center}

\subsubsection{The geometric assumption}
Let us describe the energy region where we study (\ref{eq:qper}). We
assume that $J$ is a real compact interval such that:
\begin{description}
\item[(H4a)] For all $E\in J$, the connected components of
$\Gamma_{\R}(E)$ have a non empty interior. \item[(H4b)]The
iso-energy curve $\Gamma_{\R}(E)$ contains a real branch that is an
unbounded vertical curve. \item[(H4c)] For all $E\in J$, the set
$\mathcal{W}(E)=[E-W_+,E-W_-]$ contains only isolated bands of the
periodic operator.
\end{description}
$W_+$ and $W_-$ are defined in (\ref{eq:minmax}).
\subsection{Geometric description}
\subsubsection{The set $W^{-1}(\mathbb{R})$}
As $E\in\mathbb{R}$, the set $(E-W)^{-1}(\mathbb{R})$ coincide with
$W^{-1}(\mathbb{R})$. It is $2\pi$-periodic. It consists of the real
line and of complex branches (curves) which are symmetric with
respect to the real line. There are complex branches beginning at
the real extrema of $W$ that do not cross again the real line.
\newline Consider an extremum of $W$ of order $n_i$ on the real
line, say $\zeta_i$. Near $\zeta_i$, the set $W^{-1}(\mathbb{R})$
consists of a real segment, and of $n_i-1$ complex curves symmetric
with respect to the real axis and intersecting the real axis only on
$\zeta_i$. The angle between two neighboring curves is equal to
$\displaystyle \frac{\pi}{n_i}$.\newline Let $Y>0$. We set
$S_Y=\{-Y\leq \textrm{Im}\  \zeta\leq Y\}$. We assume that $Y$ is so
small that
\begin{itemize}
\item $S_Y $ is contained in the domain of analyticity of $W$;
\item the set $W^{-1}(\mathbb{R})\cap S_Y$ consists of the real
line and of the complex lines passing through the real extrema of
$W$.
\end{itemize}
An example of subset $W^{-1}(\R)$ is shown in figure
\ref{fig:preimg}.

\psset{unit=1em,linewidth=.05}
\begin{center}
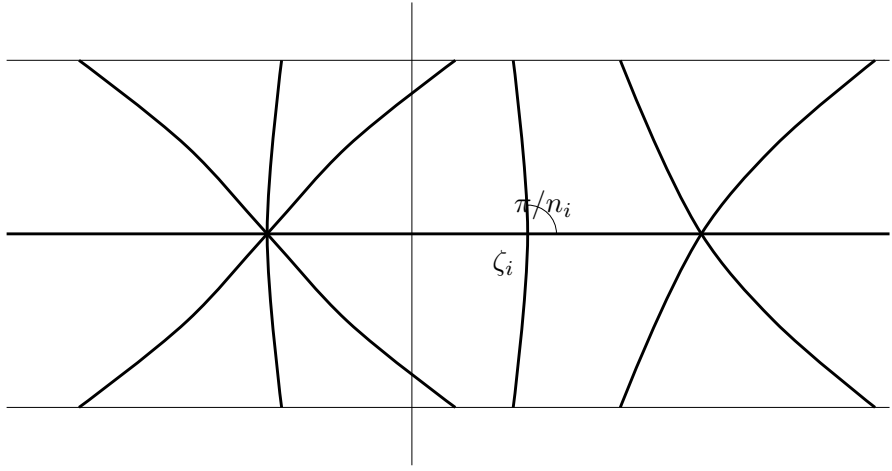
\begin{figure}
\begin{pspicture}(-20,-10)(20,10) \psline[linewidth=0.1](-14,0)(16.5,0)
\psline[linewidth=0.01](0,8)(0,-8)\psline[linewidth=0.01](-14,6)(16.5,6)
\psline[linewidth=0.01](-14,-6)(16.5,-6)
\pscurve[linewidth=0.1](-11.5,-6)(-7.75,-3)(-5,0)(-2.25,3)(1.5,6)\pscurve[linewidth=0.1](-4.5,-6)(-5,0)(-4.5,6)\pscurve[linewidth=0.1](1.5,-6)(-2.25,-3)(-5,0)(-7.75,3)(-11.5,6)
\pscurve[linewidth=0.1](3.5,-6)(4,0)(3.5,6)
\pswedge[linewidth=0.01](4,0){1}{0}{90}
\pscurve[linewidth=0.1](16,-6)(12.5,-3)(10,0)(7.2,6)\pscurve[linewidth=0.1](7.2,-6)(10,0)(12.5,3)(16,6)
\uput[180](6,1){$\pi/n_{i}$} \uput[180](4,-1){$\zeta_{i}$}
\end{pspicture}
\caption{A subset of $W^{-1}(\R)\cap S_{Y}$}\label{fig:preimg}
\end{figure}
\end{center}

\subsubsection{Notations and description of $(E-W)^{-1}(\sigma(H_{0}))$}
For all $E\in J$, we write
$$(E-W)^{-1}(\sigma(H_{0}))\cap\R=\bigcup\limits_{k\in
\Z}\bigcup\limits_{j=1}^{N}\{[\varphi_{j}^{-}(E),\varphi_{j}^{+}(E)]+2k\pi\},$$
with the following properties:
\begin{itemize}
\item
$\varphi_{1}^{-}(E)<\varphi_{1}^{+}(E)<\varphi_{2}^{-}(E)<\dots<\varphi_{N}^{-}(E)<\varphi_{N}^{+}(E),\qquad
0<\varphi_{N}^{+}(E)-\varphi_{1}^{-}(E)<2\pi.$ \item The connected
component $[\varphi_{1}^{-}(E),\varphi_{1}^{+}(E)]$ is associated to
a connected component of $\Gamma_{\R}(E)$ which is an unbounded
vertical curve. \item We generally define:
$$ \varphi_{j+k}^{-}(E)=\varphi_{j}^{-}(E)+2k\pi,\quad \forall
j\in\{1,\dots, N\},\quad\forall k\in \Z.$$
\end{itemize}
We set $\mathcal{B}_j(E)=[\varphi_j^{-}(E),\varphi_j^{+}(E)]$,
$\mathcal{G}_j(E)=]\varphi_{j}^+(E),\varphi_{j+1}^{-}(E)[$ and
$\mathcal{G}_N(E)=(\varphi^+_N(E),\varphi_1^+(E)+2\pi)=(\varphi_N^+(E),2\pi)\cup(2\pi,2\pi+\varphi_{1}^-(E))$.
Let $(\zeta_i^j)_{1\leq i\leq p_j}$ be the extrema of $W$ in
$\mathcal{G}_j$. We recall that $n_i^j$ is the order of
$\zeta_i^j$.\newline We have the following description:
\begin{Le}
\label{le:ext} Fix $[A,B]$ a compact interval of
$\R$.\smallpagebreak
 There exists a
finite number $p$ of real extrema of $W$ in $[A,B]$.
\begin{itemize}
\item If $p=0$, there exists $Y>0$ such that:
$$(E_{0}-W)^{-1}(\mathbb{R})\cap\{\zeta\in S_{Y};\
\textrm{Re}\zeta\in[A,B]\}=[A,B].$$ \item For $p>0$. We denote by
$\{\zeta_{1},\dots,\zeta_{p}\}$ the real extremum of $W$ in $[A,B]$.
There exist $Y>0$ and a sequence
$\{\Sigma_{i}^{_1},\dots,\Sigma_{i}^{n_{i}-1}\}_{i\in\{1\dots p\}}$
of disjoint strictly vertical lines of $\C_{+}$ starting at
$\zeta_{i}$ such that:
$$(E_{0}-W)^{-1}(\mathbb{R}))\cap\{\zeta\in S_{Y};\
\textrm{Re}
\zeta\in[A,B]\}=[A,B]\bigcup_{i=1}^{p}\Big(\bigcup_{k=1}^{n_{i}-1}(\Sigma_{i}^{k}\cup\overline{\Sigma_{i}^{k}})\Big).$$
\end{itemize}
\end{Le}
\subsubsection{The main result}
The main object of this paper is to prove \begin{Th}\label{latou} We
assume (H1)-(H4) are satisfied. For $\varepsilon
>0$ sufficiently small, for almost all $z\in \mathbb{R}$, one has
$$
\sigma(H_{z,\varepsilon})\cap J\neq \emptyset \ \ {\rm{and}}\ \
\sigma_{ac}(H_{z,\varepsilon})\cap J=\emptyset.
$$
Here $\sigma_{ac}(H_{z,\varepsilon})$ is the absolutely continuous
spectrum of the family of operators $(H_{z,\varepsilon})$.
\end{Th}
Using the Ishii-Pastur-Kotani Theorem \cite{quatre,PasFig} one can
see that the result of Theorem \ref{latou} is deduced from the
positivity of the Lyapunov exponent. We actually recall the
following result
\begin{Th}\cite{quatre}
Let $\Theta(E,\varepsilon)$ be the Lyapunov exponent of
(\ref{eq:qper}). We have
$$\sigma_{ac}(H_{z,\varepsilon})=\overline{\{E|\Theta(E)=0\}}^{ess}.$$
\end{Th}
\section{Periodic Schr{\"o}dinger operators} \label{sec:per} This
section is devoted to the study of the periodic Schr{\"o}dinger
operator (\ref{eq:opper}) where $V$ is a 1-periodic, real-valued,
$L^{2}_{\textrm{loc}}$- function. We recall known facts needed on
the present paper and we introduce notations. Basic references are
\cite{Eas:73, FEKL2, McK, Ti}.
\subsection{Bloch solutions}
Let $\psi$ be a solution of the equation
\begin{equation}
\label{eq:eqper}
-\frac{d^{2}}{dx^{2}}\psi(x,\mathcal{E})+V(x)\psi(x,\mathcal{E})=\mathcal{E}\psi(x,\mathcal{E}),\quad
x\in\R.
\end{equation}
satisfying the relation
\begin{equation}\label{eq:mult}
\psi(x+1,\mathcal{E})=\lambda(\mathcal{E})\psi(x,\mathcal{E}).
\end{equation}
for all $x\in\R$ and some non-vanishing complex number
$\lambda(\mathcal{E})$ independent of $x$. Such a solution exists
and is called the {\it Bloch solution} and $\lambda(\mathcal{E})$ is
called {\it Floquet multiplier}. We discuss its analytic properties
as a function of $\mathcal{E}$.\newline As in section
\ref{sec:enerregion}, we denote the spectral bands of the periodic
Schr{\"o}dinger operator by $[E_{2n+1},\,E_{2n+2}]$, $n\in\N$.
Consider $\mathcal{S}_{\pm}$ two copies of the complex plane
$\mathcal{E}\in\C$ cut along the spectral bands. Paste them together
to get a Riemann surface with square root branch points. We denote
this Riemann surface by $\mathcal{S}$.\newline One can construct a
Bloch solution $\psi(x,\mathcal{E})$ meromorphic on $\mathcal{S}$.
It is normalized by the condition $\psi(1,\mathcal{E})\equiv1$. The
poles of this solution are located in the open spectral gaps or at
their edges; the closure of each spectral gap contains exactly one
pole that, moreover, is simple. It is located either on
$\mathcal{S}_{+}$ or on $\mathcal{S}_{-}$. The position of the pole
is independent of $x$.\newline For $\mathcal{E}\in\mathcal{S}$, we
denote by $\widehat{\mathcal{E}}$ the point on $\mathcal{S}$
different from $\mathcal{E}$ and having the same projection on $\C$
as $\mathcal{E}$. We let
$$\widehat{\psi}(x,\mathcal{E})=\psi(x,\widehat{\mathcal{E}}),\quad\widehat{\mathcal{E}}\in\mathcal{S}.$$
The function $\widehat{\psi}(x,\mathcal{E})$ is another Bloch
solution of (\ref{eq:eqper}). Except at the edges of the spectrum,
the functions $\psi$ and $\widehat{\psi}$ are linearly independent
solutions of (\ref{eq:eqper}). In the spectral gaps, $\psi$ and
$\widehat{\psi}$ are real valued functions of $x$, and, on the
spectral bands, they differ only by complex conjugation.
\subsection{The Bloch quasi-momentum}
\label{sec:qmom} Consider the Bloch solution $\psi(x,\mathcal{E})$.
The corresponding Floquet multiplier $\lambda(\mathcal{E})$ is
analytic on $\mathcal{S}$. Represent it in the form
$\lambda(\mathcal{E})=\exp(i k(\mathcal{E}))$. The function
$\mathcal{E}\mapsto k(\mathcal{E})$ is the {\it Bloch
quasi-momentum} of $H_{0}$. Its inverse $k\mapsto \mathbf{E}(k)$ is
the {\it dispersion relation} of $H_{0}$. A branching point $\zeta$
is a point where $k'(\zeta)=0$. \smallpagebreak Let $D$ be a simply
connected domain containing no branch point of the Bloch
quasi-momentum. In $D$, one can fix an analytic single-valued branch
of $k$, say $k_{0}$. All the other single-valued branches of $k$
that are analytic in $D$ are related to $k_{0}$ by the formulae
\begin{equation}
\label{eq:qmdet} k(\mathcal{E})=\pm k_{0}(\mathcal{E})+2\pi l,\quad
l\in\Z.
\end{equation}
Consider $\C_{+}$, the upper half plane of the complex plane. On
$\C_{+}$, one can fix a single valued analytic branch of the
quasi-momentum continuous up to the real line. It can be determined
uniquely by the conditions $\textrm{Re}\ k(\mathcal{E}+i0)=0$ and
$\textrm{Im}\ k(\mathcal{E}+i0)>0$ for $\mathcal{E}<E_{1}$. We call
this branch the main branch of the Bloch quasi-momentum and denote
it by $k_{p}$.\smallpagebreak The function $k_{p}$ conformally maps
$\C_{+}$ onto the first quadrant of the complex plane cut at compact
vertical slits starting at the points $\pi l,\ l\in\N$. It is
monotonically increasing along the spectral zones so that
$[E_{2n-1},E_{2n}]$, the n-th spectral band, is mapped on the
interval $[\pi(n-1),\pi n]$. Along any open gap, $\textrm{Re}\
k_{p}(\mathcal{E}+i0)$ is constant, and $\textrm{Im}\
k_{p}(\mathcal{E}+i0)$ is positive and has only one non-degenerate
maximum. \smallpagebreak All the branch point of $k_{p}$ are of
square root type. Let $E_{l}$ be a branch point of $k_p$. In a
sufficiently small neighborhood of $E_{l}$, the function $k_{p}$ is
analytic in $\sqrt{E-E_{l}}$, and
\begin{equation}
\label{eq:qmsqrt}
k_{p}(\mathcal{E})-k_{p}(E_{l})=c_{l}\sqrt{\mathcal{E}-E_{l}}+o(\mathcal{E}-E_{l})\quad
c_{l}\neq 0.
\end{equation}
Finally, we note that the main branch can be continued analytically
to the complex plane cut along $(-\infty,E_{1}]$ and the spectral
gaps $]E_{2n},E_{2n+1}[$, $n\in\N^{*}$, of the periodic operator
$H_{0}$.
\subsection{A meromorphic function}
\label{sec:omega} Now let us discuss a function playing an important
role in the adiabatic constructions.\smallpagebreak In \cite{FEKL2},
it is shown that, on $\mathcal{S}$, there is a meromorphic function
$\omega$ having the following properties:
\begin{itemize}
\item the differential $\Omega=\omega d\mathcal{E}$ is
meromorphic; its poles are the points of $P\cup Q$, where $P$ is the
set of poles of $\mathcal{E}\mapsto\psi(x,\mathcal{E})$, and $Q$ is
the set of zeros of $k'$; \item all the poles of $\Omega$ are
simple; \item if the residue of $\Omega$ at a point $p$ is denoted
by $\textrm{res}_{p}\Omega$, one has
\begin{eqnarray}
\textrm{res}_{p}\Omega=1,\forall p\in P\setminus Q,\\
\textrm{res}_{q}\Omega=-1/2,\forall q\in Q\setminus P,\\
\textrm{res}_{r}\Omega=1/2,\forall r\in P\cap Q.
\end{eqnarray}
\item if $\mathcal{E}\in\mathcal{S}$ projects into a gap, then
$\omega(\mathcal{E})\in\R$. \item if $\mathcal{E}\in\mathcal{S}$
projects inside a band, then
$\overline{\omega(\mathcal{E})}=\omega(\widehat{\mathcal{E}})$.
\end{itemize}
\subsubsection{The complex momentum}
It is the main analytic object of the complex WKB method. Let
$\zeta\in S_{Y}$. We define $\kappa$, in $\mathcal{D}(W)$ the domain
of analyticity of $W$ by  by :
\begin{equation}\label{eq:comp}
\kappa(\zeta)=k(E-W(\zeta)).
\end{equation}
Here, $k$ is the Bloch quasi-momentum defined in section
\ref{sec:qmom}. Though $\kappa$ depends on $E$, we omit the
$E$-dependence. Relation (\ref{eq:comp}) translates the properties
of $k$ into properties of $\kappa$. Hence,
$\zeta\mapsto\kappa(\zeta)$ is a multi-valued analytic function, and
its branch points are related to the branch points of the
quasi-momentum by the relations
\begin{equation}\label{eq:branchpoints}
E-W(\zeta)=E_{l},\quad l=1,2,3,\cdots
\end{equation}
Let $\zeta_{0}$ be a branch point of $\kappa$. If $W'(\zeta_{0})\neq
0$, then $\zeta_{0}$ is a branch point of square root type.\newline
If $D\subset D(W)$ is a simply connected set containing no branch
points of $\kappa$, we call it {\it regular}. Let $\kappa_{p}$ be a
branch of the complex momentum analytic in a regular domain $D$. All
the other branches that are analytic in $D$ are described by the
following formulae:
\begin{equation}\label{eq:comprel}
\kappa_{m}^{\pm}=\pm\kappa_{p}+2\pi m.
\end{equation}
Here $\pm$ and $m\in \mathbb{Z}$ are indexing the branches.
\subsubsection{Index of an interval $[\varphi_{j}^{-}(E),\varphi_{j}^{+}(E)]$}
\label{sec:ind} Fix $j\in\{1,\dots,N\}$. Fix a continuous branch
$\kappa_{j}$ of the complex momentum on
$[\varphi_{j}^{-}(E),\varphi_{j}^{+}(E)]$. We define:
\begin{equation}
\label{eq:ind} \kappa_{j}^{+}=\kappa_{j}(\varphi_{j}^{+});\quad
\kappa_{j}^{-}=\kappa_{j}(\varphi_{j}^{-});\quad
p_{j}\pi=\kappa_{j}^{+}-\kappa_{j}^{-}.
\end{equation}
$p_{j}$ is the {\it index} of
$[\varphi_{j}^{-}(E),\varphi_{j}^{+}(E)]$ associated to
$\kappa_{j}$.\smallpagebreak Let us give some properties of $p_{j}$.
\begin{Le}\label{le:ind}
Assume that $(H4)$ is satisfied. The indexes $p_{j}$ have the
following properties:
\begin{enumerate}
\item For $j\in\{1,\cdots,N\}$, $p_{j}\in\{-1,0,1\}$. \item
$\sum_{j=1}^{N}|p_{j}|\in 2\N.$
\end{enumerate}
\end{Le}
{\bf{Proof}} The points $(E-W)(\varphi_{j}^{-}(E))$ and
$(E-W)(\varphi_{j}^{+}(E))$ are the ends of a band of
$\sigma(H_{0})$: they are distinct or coincide. If they coincide,
i.e. if $(E-W)(\varphi_{j}^{-}(E))=(E-W)(\varphi_{j}^{+}(E))$, the
index $p_{j}$ satisfies $p_{j}=0$. Else, we consider $k_{j}$ the
branch of the quasi-momentum associated to $\kappa_{j}$. We have
that
$$|k_{j}((E-W)(\varphi_{j}^{-}(E)))-k_{j}((E-W)(\varphi_{j}^{+}(E)))|=\pi,$$
see (\ref{sec:qmom}), and $|p_{j}|=1$.\smallpagebreak Let us prove
point (2). We write:
$$\sum_{j=1}^{N}|p_{j}|\equiv\sum_{j=1}^{N}p_{j}\ [2],$$
$$\sum_{j=1}^{N}p_{j}=\sum_{j=1}^{N}\frac{\kappa_{j}(\varphi_{j}^{+})-\kappa_{j}(\varphi_{j}^{-})}{\pi}=\frac{\kappa_{N}(\varphi_{N}^{+})-\kappa_{1}(\varphi_{1}^{-})}{\pi}+\sum_{j=1}^{N-1}\frac{\kappa_{j+1}(\varphi_{j+1}^{-})-\kappa_{j}(\varphi_{j}^{+})}{\pi}.$$
For $j\in\{1,\cdots,N-1\}$, $(E-W)(\varphi_{j+1}^{-}(E))$ and
$(E-W)(\varphi_{j}^{+}(E))$ are the ends of a same gap and
$$k_{j+1}((E-W)(\varphi_{j+1}^{-}(E)))\equiv k_{j}((E-W)(\varphi_{j}^{+}(E)))[2\pi].$$
By periodicity, $(E-W)(\varphi_{1}^{-}(E))$ and
$(E-W)(\varphi_{N}^{+}(E))$ are the ends of the same
gap.\smallpagebreak This ends the proof of Lemma
\ref{le:ind}.\smallpagebreak If $p_{j}\neq 0$, we say that we cross
a band. In this case, $(E-W)(\varphi_{j}^{-})\neq
(E-W)(\varphi_{j}^{+})$ and the associated connected component of
the iso-energy curve is unbounded vertically.\smallpagebreak We
notice that $p_{1}\neq 0$, and thus $\sum_{j=1}^{N}|p_{j}|>0$.
\begin{Remark}
We can shose the determination of $\mathcal{\kappa}$ such that
$p_1=1$.
\end{Remark}
\subsubsection{Tunneling coefficients}
For $j\in\{1,\dots,N\}$, we denote by $\gamma_{j}$ a smooth closed
curve that goes once around
$[\varphi_{j}^{+}(E),\varphi_{j+1}^{-}(E)]$. Notice that this curve
is the projection of a closed curve on the complex Riemann surface
$\kappa(\zeta)=k(E-W(\zeta))$. We consider the tunneling actions
$S_{j}$ given by:
\begin{equation}
\label{eq:act} S_{j}(E)=i\oint_{\gamma_{j}}\kappa(\zeta)d\zeta,\
\quad \forall j\in\{1,\dots N\},
\end{equation}
It is straightforward to prove that for $E\in J$, each of these
actions is real and non-zero and that $S_j$ is analytic in a complex
neighborhood of $J$ (for analogous statements, we refer to
\cite{FK0, Mg1}). By definition, we choose the direction of the
integration so that all the tunneling actions be positive. We set
$$t_j=e^{-\frac{1}{2\varepsilon}S_j}.$$
\begin{equation}
\label{eq:tunnel}
T(E,\varepsilon)=e^{-\sum\limits_{j=1}^{N}S_{j}(E)/2\varepsilon}.
\end{equation}
$T$ is exponentially small.
\section{The Proof of Theorem \ref{latou}}\label{lili}
\subsubsection{Spectral results}
One of the main objects of the spectral theory of quasi-periodic
operators is the Lyapunov exponent (for a definition and additional
information, see, for example, \cite{quatre}). The main result of
this section is
\begin{Th}
\label{th:lyap} We assume that the assumptions $(H1)-(H4)$ are
satisfied. Then, on the interval $J$, for sufficiently small
irrational $\varepsilon/2\pi$, the Lyapunov exponent
$\Theta(E,\varepsilon)$ of (\ref{eq:qper}) is positive and satisfies
the asymptotics
\begin{equation}\label{eq:lyap}
\Theta(E,\varepsilon)=\frac{\varepsilon}{2\pi}\sum_{j=1}^N\ln\frac{1}{t_j}+o(1)=\frac{1}{4\pi}
\sum\limits_{j=1}^{N}S_{j}(E)+o(1).
\end{equation}
\end{Th}
This theorem implies that if $\varepsilon/2\pi$ is sufficiently
small and irrational, then, the Lyapunov exponent is positive for
all $E\in J$.
\subsection{The monodromy matrix and the Lyapunov exponents}
The main object of our study in this subsection is the monodromy
matrix for the family of equations (\ref{eq:qper}), we define it
briefly (we refer the reader to \cite{FK0,FEKL1}). In this paper, we
compute the asymptotics of its Fourier expansion in the adiabatic
limit.
\subsubsection{Definition of the monodromy
matrix}\label{sec:monmat} Fix $E\in\R$. Consider the family of
differential equations indexed by $z\in\R$,
\begin{equation}\label{eq:dec}
\left(-\frac{d^{2}}{dx^{2}}+V(x-z)+W(\varepsilon
x)\right)\psi(x)=E\psi(x).
\end{equation}
\begin{Def}
We say that $(\psi_{i})_{i\in\{1,2\}}$ is a consistent basis of
solutions to (\ref{eq:dec}) if the two functions $((x,z)\mapsto
\psi_{i}(x,z,E))_{i\in\{1,2\}}$ are a basis of solutions to
(\ref{eq:dec}) whose Wronskian is independent of $z$ and that are
1-periodic in $z$ i.e that satisfy
\begin{equation}\label{eq:consbis}
\forall x\in\R, \quad\forall z\in\R \quad,\forall i\in\{1,2\},\quad
\psi_{i}(x,z+1,E)=\psi_{i}(x,z,E).
\end{equation}
\end{Def}
We refer the reader to \cite{FEKL1,FEKL2} about the existence and
details on consistent basis of solutions to (\ref{eq:dec}) .\newline
 The functions $((x,z)\mapsto
\psi_{i}(x+2\pi/\varepsilon,z+2\pi/\varepsilon,E))_{i\in\{1,2\}}$
being also solutions of equation (\ref{eq:dec}), we get the relation
\begin{equation}\label{eq:mondef}
\Psi(x+2\pi/\varepsilon,z+2\pi/\varepsilon,E)=M(z,E)\Psi(x,z,E),
\end{equation}
where
\begin{itemize}
\item $\Psi(x,z,E)=(\begin{array}c \psi_{1}(x,z,E)\\
\psi_{2}(x,z,E)\end{array}),$ \item $M(z,E)$ is a $2\times 2$-matrix
with coefficients independent of $x$.
\end{itemize}
The matrix $M$ is called the {\it monodromy matrix} associated to
the consistent basis $(\psi_{1,2})$. We recall the following
properties of this matrix.
\begin{equation}
\det M(z,E)\equiv 1,\quad M(z+1,E)=M(z,E),\quad \forall z\in
\mathbb{R}.
\end{equation}
The Matrix $M$ belongs to $SL(2,\mathbb{R})$ which is known to be
isomorph to $SU(1,1)$.
\subsection{The Lyapunov exponents and the monodromy equation}
Consider now a 1-periodic, $SL(2,\C)$-valued function, say,
$z\mapsto \widetilde{M}$, and $h>0$ irrational. Consider the finite
difference equation
\begin{equation}
F_{n+1}=\widetilde{M}(z+nh)F_{n}\quad \forall n\in\Z, \quad
F_{n}\in\C^{2}.\label{eq:finid}
\end{equation}
Going from equation (\ref{eq:qper}) to the (\ref{eq:finid}) is close
to the monodromization  transformation introduced in \cite{huit} to
construct Bloch solutions of difference equation. Indeed, it appears
that the behavior of solutions of (\ref{eq:qper}) for
$x\rightarrow\mp \infty$ repeats the behavior of solutions of the
monodromy equation for $n\rightarrow\mp \infty$. And it is a well
known fact that the spectral properties of the one dimensional
Schr\"odinger equations can be described in terms of the behavior of
its solutions as $x\rightarrow \mp \infty $.\newline The Lyapunov
exponent of the finite difference equation (\ref{eq:finid}) is
\begin{equation}\label{eq:coc}
\theta(\widetilde{M},h)=\lim\limits_{N\mapsto
+\infty}\frac{1}{N}\log\|P_{N}(z,h)\|.
\end{equation}
where the matrix cocycle $(P_{N}(z,h))_{N\in\N}$ is defined as
\begin{equation}
P_{N}(z,h)=\widetilde{M}(z+Nh)\cdot \widetilde{M}(z+(N-1)h)\cdots
\widetilde{M}(z+h)\cdot \widetilde{M}(z).
\end{equation}
It is well known that, if $h$ is irrational, and $\widetilde{M}$ is
sufficiently regular in $z$, then the limit (\ref{eq:coc}) exists
for almost all $z$ and is independent of $z$. \newline Set $h\equiv
2\pi/\varepsilon\ [1]$. Let $M$ be the monodromy matrix associated
to a consistent basis $(\psi_{1,2})$. Consider the {\it monodromy
equation}
\begin{equation}\label{eq:moneq}
F_{n+1}=M(z+nh,E)F_{n}\quad \forall n\in\Z, \quad F_{n}\in\C^{2}.
\end{equation}
The Lyapunov exponent of the monodromy equation (\ref{eq:moneq}) is
defined by
$$ \theta(E,\varepsilon)=\theta(M(z,E),h).$$
There are several deep relations between equation (\ref{eq:dec}) and
the monodromy equation (\ref{eq:moneq}) (see \cite{FEKL1,Fe}). We
describe only one of them. Recall that $ \Theta(E,\varepsilon)$ is
the Lyapunov exponent of the equation (\ref{eq:qper}) We have the
following result.
\begin{Th}(\cite{FEKL1})Assume $\varepsilon/2\pi$ is irrational. The Lyapunov
exponent $ \Theta(E,\varepsilon)$ and $ \theta(E,\varepsilon)$ are
related by the following relation.
\begin{equation}\label{eq:lyaprel}
\Theta(E,\varepsilon)=\frac{\varepsilon}{2\pi}\theta(E,\varepsilon).
\end{equation}
\end{Th}
\subsection{The asymptotics of the monodromy matrix}
As $W$ and $V$ are real on the real line, we construct a monodromy
matrix of the form
\begin{equation}
M(z,E)=\left(\begin{array}{cc}A(z,E)&B(z,E)\\
\overline{B(\overline{z},\overline{E})}&\overline{A(\overline{z},\overline{E})}\end{array}\right)
\label{latou2}
\end{equation}
To get (\ref{latou2}), it suffices to consider a basis of solutions
of the form $(u;\overline{u})$. For the details on the existence and
the construction of such a basis are developed in
\cite{FK0}.\\
 The following result
gives the asymptotics of $A$ and $B$ in the adiabatic case.
\begin{Th}\label{latouf2}
Let $E_{0}$ be in $J$. There exists $Y>0$ and $V_{0}$, a
neighborhood of $E_{0}$, such that, for sufficiently small
$\varepsilon$, the family of equations (\ref{eq:dec}) has a
consistent basis of solutions for which the corresponding monodromy
matrix $M$ is analytic in $(z,E)\in\{z\in\C\ ;\ |\textrm{Im}
z|<Y/\varepsilon\}\times V_{0}$ and has the form (\ref{latou2}).
When $\varepsilon$ tends to $0$, the coefficients $A$ and $B$ admit
the asymptotics
\begin{equation}\label{eq:moncoeffplus}
A=A_{+}(E,\varepsilon)e^{-2 iQ_+\pi z}[1+o(1)],\quad
B=B_{+}(E,\varepsilon)e^{-2 iP_+\pi z}[1+o(1)],\quad 0<\textrm{Im}
z<Y/\varepsilon,
\end{equation}
\begin{equation}\label{eq:moncoeffmoins}
A=A_{-}(E,\varepsilon)e^{2 iQ_-\pi z}[1+o(1)],\quad
B=B_{-}(E,\varepsilon) e^{2 iP_- \pi z}[1+o(1)],\quad
-Y/\varepsilon<\textrm{Im} z<0.
\end{equation}
The integers $P_-,P_+$ and $Q_-,Q_+\in\Z$ are specified in section
\ref{lili}. There exists a constant $C>1$ such that for
$\varepsilon>0$ sufficiently small and $E\in V_{0}\cap\R$, one has
\begin{equation}\label{eq:moncoeffenc}
\frac{1}{C}<T(E,\varepsilon)|A_{\pm}(E,\varepsilon)|<C,\quad
\frac{1}{C}<T(E,\varepsilon)|B_{\pm}(E,\varepsilon)|<C,
\end{equation}
where $T(E,\varepsilon)$ is defined in (\ref{eq:tunnel}). \newline
For $Y_{1}$ and $Y_{2}$ such that $0<Y_{1}<Y_{2}<Y$, there exists
$V=V(Y_{1},Y_{2})$ a neighborhood of $E_{0}$ such that the
asymptotics (\ref{eq:moncoeffplus}) and (\ref{eq:moncoeffmoins}) for
$A$ and $B$ are uniform in $(z,E)\in\{z\in\C\ ;\
Y_{1}/\varepsilon<|\textrm{Im} z|<Y_{2}/\varepsilon\}\times V$.
\end{Th}
\begin{Remark}
The coefficients $A_+$, $A_-,\ B_+$ and $B_-$ are the leading terms
of the asymptotics of the $Q_{\mp}$-th and $P_{\mp}$-th Fourier
coefficients of the monodromy matrix coefficients. From Theorem
\ref{latouf2}, one deduces that, in the strip $\{-Y<\textrm{Im}
\zeta<Y\}$, only a few Fourier series terms of the monodromy matrix
dominate.
\end{Remark}
The proof of Theorem \ref{latou} is given in section \ref{lili}.
\subsection{The asymptotics for the Lyapunov exponent}
\subsubsection{The upper bound}
Fix $P=\max{|P_{\mp}|,|Q_{\mp}|}$. The asymptotics
(\ref{eq:moncoeffplus}) and (\ref{eq:moncoeffmoins}) and estimates
(\ref{eq:moncoeffenc}) imply the following estimates for the
coefficients of $M(z,E)$, the monodromy matrix:
\begin{equation}
|A|,|B|\leq C(y_0)\cdot T(E)^{-1}e^{2\pi P y_0/\varepsilon},\
\textrm{Im}z=y_0/\varepsilon,\label{simou1}
\end{equation}
\begin{equation} |A|,|B|\leq C(y_0)\cdot T(E)^{-1}e^{2\pi
P y_0/\varepsilon},\
\textrm{Im}z=-y_0/\varepsilon.\label{simou2}\end{equation} Here,
$C(y_0)$ is a positive constant independent of $\varepsilon$,
$\textrm{Re}z$, and $E$. The estimates are valid for sufficiently
small $\varepsilon$. We recall that $M$ is analytic and $1$-periodic
in $z$. Equations (\ref{simou1}),  (\ref{simou2}) and the maximum
principle imply that:
$$
|A|,|B|\leq 2C(y_0)T(E)^{-1}\exp(2\pi Py_0/\varepsilon),\ \ \ z\in
\mathbb{R}.
$$
This leads to the following upper bound for the Lyapunov exponent
for the matrix cocycle generated by $M(z,E)$.
\begin{equation}
\theta(E,\varepsilon)\leq \log(T(E)^{-1})+C+2\pi P y_0/\varepsilon;
\end{equation}
where $C$ is a constant independent of $E$ and $\varepsilon$. Using
(\ref{eq:lyaprel}) one gets
\begin{equation}
\Theta(E,\varepsilon)\leq \frac{\varepsilon}{2\pi}\log
(T(E)^{-1})+\varepsilon C+2\pi Py_0.\label{lot1}
\end{equation}
\subsubsection{The lower bound} For
$(M(z,\varepsilon)_{0<\varepsilon<1})$ a family of
 $SL(2,\mathbb{C})$-valued, $1$-periodic functions of $z\in
 \mathbb{C}$ and $h$ an irrational number, we recall the following
 result obtained in \cite{huit}.
 \begin{Pro}\label{youswa}
 Fix $\varepsilon_0 >0$. Assume that there exist $y_0$ and $y_1$
 such that $0<y_0<y_1<\infty$ and such that, for any $\varepsilon
 \in (0,\varepsilon_0) $ one has
 \begin{itemize}
 \item the function $z\to M(z,\varepsilon)$ is analytic in the
 strip $S=\{z\in \mathbb{C};\ 0\leq \textrm{Im}z\leq
 \frac{y_1}{\varepsilon}\}$;
 \item in the strip $S_1=\{z\in \mathbb{C};\
 \frac{y_0}{\varepsilon}\leq \textrm{Im}z \leq
 \frac{y_1}{\varepsilon}\}\subset S, M(z,\varepsilon)$ admits the
 representation
 \begin{equation}
 M(z,\varepsilon)=\lambda(\varepsilon)e^{i2\pi n_0z}\cdot
 \big(M_0(\varepsilon)+M_1(z,\varepsilon)\big);
 \end{equation}
 for some constant $\lambda(\varepsilon)$, some integer $n_0$ and
 a matrix $M_0(\varepsilon)$, all of them independent of $z$;
 \item $
M(z,E)=\left(\begin{array}{cc}1&0\\
\beta(\varepsilon)&\alpha(\varepsilon)\end{array}\right)$; \item
there exist constants $\beta>0$ and $\alpha\in (0,1)$ independent of
$\varepsilon$ such that $\mid \alpha(\varepsilon)\mid \leq \alpha$
and $\mid \beta(\varepsilon) \mid\leq \beta $; \item
$m(\varepsilon)=\sup_{z\in S_1}\parallel M_1(z,\varepsilon)\parallel
\rightarrow 0$ as $\varepsilon \rightarrow 0$.
 \end{itemize}
 Then, there exists $C>0$ and $\varepsilon_1>0$ (both depending
 only on $y_0,\ y_1,\ \alpha,\ \beta$ and $\varepsilon \mapsto
 m(\varepsilon)) $ such that, if $0<\varepsilon<\varepsilon_1$,
 one has
 \begin{equation}\label{naj1}
 \theta(M(\cdot,\varepsilon),h)>\log|\lambda(\varepsilon)|-Cm(\varepsilon)
 \end{equation}
 \end{Pro} Proposition \ref{youswa} is used by
 applying the arguments of \cite{FEKL1,FEKL2} to get the lower
bound for the Lyapunov exponent. \\ First for $ \sigma
=\left(\begin{array}{cc} 0& 1\\ 1&0\end{array}\right)$ we prove that
the matrix $\sigma M(z,E)\sigma $ completes the assumption of
Proposition \ref{youswa}. \\ Let $y_0$ and $y_1$ fixed such that
$0<y_0<y_1<Y$. The asymptotics of the monodromy matrix are uniform
for $z$ in $S=\{z\in\mathbb{C};\ y_0/\varepsilon \leq \textrm{Im}
z\leq y_1/\varepsilon\}$ and $E\in V_0$. \newline Let us assume that
$n=N_N$ in (\ref{recto}) is even then the following relations holds
\begin{equation} Q_-=Q_++1, \ Q_-=P_-+1\  \rm{and}\ Q_-=P_+
\end{equation}For $E\in V_0\cap \mathbb{R}$ and $z\in S$, Theorem
\ref{latouf2} implies that
$$
\overline{A(\overline{z},E)}=\overline{A_-(E,\varepsilon)}e^{-2\pi
iQ_- z}(1+o(1)),\ \
\frac{A(z,E)}{\overline{A(\overline{z},E)}}=o(1),
$$
$$
\frac{\overline{B(\overline{z},E)}}{\overline{A(\overline{z},E)}}=o(1),\
\ \frac{B(z,E)}{\overline{A(\overline{z},E)}}=c(E)(1+o(1)),
$$
Where $c(E)$ is independent of $z$ and bounded by a constant
uniformly in $\varepsilon$ and $E$. So, we have
$$\sigma \cdot M(z,E)\cdot \sigma=\overline{A_-(E,\varepsilon)}e^{-2\pi i
Q_-z}\cdot \Big( \left(\begin{array} {cc}1&0 \\
c(E)&0\end{array}\right) +o(1)\Big).
$$
This gives that the matrix-valued function $$z\mapsto \sigma \cdot
M(z,E)\cdot \sigma ;$$ satisfies the assumptions of Proposition
\ref{youswa}. \begin{Remark} When $n=N_N$ in (\ref{recto}) is odd
then we get
\begin{equation}
Q_+=Q_-+1,\ P_-=Q_+, \ \ {\rm{and}} P_++1=Q_+
\end{equation}For $E\in V_0\cap \mathbb{R}$ and $z\in S$, Theorem
\ref{latouf2} implies that
$$
A(z,E)=A_+(E,\varepsilon)e^{-2\pi iQ_+z}(1+o(1)),\ \
\frac{\overline{A(\overline{z},E)}}{A(z,E)}=o(1),
$$
$$
\frac{\overline{B(\overline{z},E)}}{A(\overline{z},E)}=c(E)(1+o(1)),\
\ \frac{B(z,E)}{A(z,E)}=o(1),
$$
Where $c(E)$ is independent of $z$ and bounded by a constant
uniformly in $\varepsilon$ and $E$. So, we have
$$\sigma \cdot M(z,E)\cdot \sigma=A_+(E,\varepsilon)e^{-2\pi i
Q_+z}\cdot \Big( \left(\begin{array} {cc}0&c(E) \\
0&1\end{array}\right) +o(1)\Big).$$ The moste important properties
of the matrix $M(z,E)$ in Proposition \ref{youswa}, is that the
bigger eigenvalue is $1$ \cite{huit}. \end{Remark}Using
(\ref{naj1}), we get that the Lyapunov exponent $\theta
(E,\varepsilon) $ of the matrix cocycle associated to
$(M(\cdot,E),h)$, satisfies the estimates
\begin{equation}
\theta(E,\varepsilon)\geq \log |A_-|+o(1).
\end{equation}
Taking into account (\ref{latou}) we get that
$$
\Theta(E,\varepsilon)\geq \frac{\varepsilon}{2\pi}\log
|A_-|+o(\varepsilon).
$$
By (\ref{eq:moncoeffenc}) we get
\begin{equation}
\Theta(E,\varepsilon)\geq \frac{\varepsilon}{2\pi}\log
(T(E)^{-1})+O(\varepsilon)\label{lot2}
\end{equation}
\subsubsection{Conclusion}
 Now we obtain (\ref{eq:lyap}), by comparing (\ref{lot1}) and
(\ref{lot2}). Indeed we see that
\begin{equation}
\Theta(E,\varepsilon)=\frac{\varepsilon}{2\pi}\log(T(E)^{-1})+o(1).\label{lot3}
\end{equation}
The expression of $T(E)$ given by (\ref{lot3}) and (\ref{lot2}) give
(\ref{eq:lyap}) for any $E\in V_0\cap\mathbb{R}$.\newline Recall
that $V_0\cap \mathbb{R}$ is an open interval containing $E_0\in J$.
The above construction can be carried out for any $E_0\in J$. The
end of the proof of Theorem \ref{latou} follows from the compactness
of the interval $J$.
 \section{The complex WKB method for adiabatic problems}
\label{sec:wkb} In this section, following \cite{FEKL2, neuf, KM1},
we describe the complex WKB method for adiabatically perturbed
periodic Schr{\"o}dinger equations
\begin{equation}\label{eq:qperbis}
-\frac{d^2}{dx^2}\psi(x)+[V(x)+W(\varepsilon
x+\zeta)]\psi(x)=E\psi(x),\quad x\in\R.
\end{equation}
Here, $V$ is 1-periodic and real valued, $\varepsilon$ is a small
positive parameter, and the energy $E$ is complex; one assumes that
$V$ is $L^{2}_{\textrm{loc}}$ and that $W$ is analytic in a strip in
the neighborhood $S_{Y}$ of the real line. \smallpagebreak The
parameter $\zeta$ is an auxiliary complex parameter used to decouple
the slow variable $\zeta=\varepsilon x$ and the fast variable $x$.
The idea of this method is to study solutions of (\ref{eq:qperbis})
in some domains of the complex plane of $\zeta$ and, then to recover
information on their behavior in $x\in\R$. Therefore, for $D$ a
complex domain, one studies solutions satisfying the condition:
\begin{equation}\label{eq:coh}
\psi(x+1,\zeta)=\psi(x,\zeta+\varepsilon)\quad\forall\zeta\in D.
\end{equation}
The aim of the WKB method is to construct solutions to
(\ref{eq:qperbis}) satisfying (\ref{eq:coh}) and that have simple
asymptotic behavior when $\varepsilon$ tends to $0$. This is
possible in certain special domains of the complex plane of $\zeta$.
These domains will depend continuously on $V,W$ and $E$. We shall
use these solutions to compute the monodromy matrix; we consider $V$
and $W$ as fixed and construct the WKB objects and solutions in an
uniform way for energies in a neighborhood of $E$.
\subsection{Standard behavior of consistent solutions}
We start by defining another analytic object central to the complex
WKB method, the {\it canonical Bloch solutions}. Then, we describe
the {\it standard behavior} of the solutions.
\subsubsection{Canonical Bloch solutions}
To describe the asymptotic formulae of the complex WKB method, one
needs to construct Bloch solutions to the equation
\begin{equation}\label{eq:perdec}
-\frac{d^{2}}{dx^{2}}\psi(x)+V(x)\psi(x)=\mathcal{E}(\zeta)\psi(x),\quad
\mathcal{E}(\zeta)=E-W(\zeta),\quad x\in\R;
\end{equation}
that are moreover analytic in $\zeta$ on a given regular
domain.\smallpagebreak Let $\zeta_{0}$ be a regular point (i.e.
$\zeta_0$ is not a branch point of $\kappa$). Let
$\mathcal{E}_{0}=\mathcal{E}(\zeta_{0})$. Assume that
$\mathcal{E}_{0}\notin P\cup Q$. Let $U_{0}$ be a sufficiently small
neighborhood of $\mathcal{E}_{0}$, and let $V_{0}$ be a neighborhood
of $\zeta_{0}$ such that $\mathcal{E}(V_{0})\subset U_{0}$. In
$U_{0}$, we fix a branch of the function $\sqrt{k'(\mathcal{E})}$
and consider $\psi_{\pm}(x,\mathcal{E})$, the two branches of the
Bloch solution $\psi(x,\mathcal{E})$ and $\Omega_{\pm}$, the
corresponding branches of $\Omega$ (see section \ref{sec:omega}. For
$\zeta\in V_{0}$, we set
\begin{equation}\label{eq:blochsol}
\Psi_{\pm}(x,\zeta)=q(\mathcal{E})e^{\int_{\mathcal{E}_{0}}^{\mathcal{E}}\Omega_{\pm}}\psi_{\pm}(x,\mathcal{E}),\quad
q(\mathcal{E})=\sqrt{k'(\mathcal{E})}, \quad
\mathcal{E}=\mathcal{E}(\zeta).
\end{equation}
The functions $\Psi_{\pm}$ are called the {\it canonical Bloch
solutions normalized at the point $\zeta_{0}$}.\smallpagebreak The
properties of the differential $\Omega$ imply that the solutions
$\Psi_{\pm}$ can be analytically continued from $V_{0}$ to any
regular domain containing $V_{0}$.\newline The Wronskian of
$\Psi_{\pm}$ satisfies (see \cite{FK0})
\begin{equation}\label{eq:wronskblochsol}
w(\Psi_{+}(\cdot,\zeta),\Psi_{-}(\cdot,\zeta))=w(\Psi_{+}
(\cdot,\zeta_{0}),\Psi_{-}(\cdot,\zeta_{0}))=k'(\mathcal{E}_{0})w(\psi_{+}(\cdot,\zeta_{0}),\psi_{-}(\cdot,\zeta_{0})).
\end{equation}
For $\mathcal{E}_{0}\notin Q\cup\{E_{l}\}$, the Wronskian
$w(\Psi_{+}(\cdot,\zeta),\Psi_{-}(\cdot,\zeta))$ is non-zero.
\subsection{Solutions having standard asymptotic behavior}
Fix $E=E_{0}$. Let $D$ be a regular domain ( i.e $D\subset D(W)$,
and simply connected set containing no branch points of $\kappa$.) .
Fix $\zeta_{0}\in D$ so that $\mathcal{E}(\zeta_{0})\notin P\cup Q$.
Let $\kappa$ be a continuous branch of the complex momentum in $D$,
and let $\Psi_{\pm}$ be the canonical Bloch solutions normalized at
$\zeta_{0}$ defined on $D$ and indexed so that $\kappa$ be the
quasi-momentum for $\Psi_{+}$.
\begin{Def}
Let $\sigma\in\{+,-\}$. We say that, in $D$, a consistent solution
$f$ has standard behavior (or standard asymptotics) if
\begin{itemize}
\item there exists $V_{0}$, a complex neighborhood of $E_{0}$, and
$X>0$ such that $f$ is defined and satisfies (\ref{eq:qperbis}) and
(\ref{eq:coh}) for any $(x,\zeta,E)\in (-X,X)\times D \times V_{0}$;
\item $f$ is analytic in $\zeta\in D$ and in $E\in V_{0}$;
\item for any compact set $K\subset D$, there exists $V\subset
V_{0}$, a neighborhood of $E_{0}$, such that, for $(x,\zeta,E)\in
(-X,X)\times K \times V$, $f$ has the uniform asymptotic
\begin{equation}\label{eq:asstd}
f=e^{\sigma\frac{i}{\varepsilon}\int_{\zeta_{0}}^{\zeta}\kappa(u)du}(\Psi_{\sigma}+o(1))\
\textrm{as $\varepsilon$ tends to $0$};
\end{equation}
\item this asymptotic can be differentiated once in $x$ without
loosing its uniformity properties.
\end{itemize}
We set \begin{equation}
f^*(x,\zeta,E,\varepsilon)=\overline{f(x,\overline{\zeta},\overline{E},\varepsilon)}
\end{equation}
\end{Def}
We call $\zeta_{0}$ the normalization point for $f$. To say that a
consistent solution $f$ has standard behavior, we will use the
following notation
$$f\sim \exp\left(\sigma
\frac{i}{\varepsilon}\int_{\zeta_{0}}^{\zeta}\kappa(u)du\right)\Psi_{\sigma}.$$
\subsection{Some results on the continuation of asymptotics}
\subsubsection{Description of the Stokes lines near
$[\varphi_{1}^{-}(E),\varphi_{1}^{+}(E)]$} This section is devoted
to the description of the Stokes lines under assumption (H4).
\subsubsection{Definition} The definition of the Stokes lines is
fairly standard, \cite{Fe, FK0}. The integral
$\zeta\mapsto\int^{\zeta}\kappa(u)du$ has the same branch points as
the complex momentum. Let $\zeta_{0}$ be one of them. Consider the
curves beginning at $\zeta_{0}$, and described by the equation
\begin{equation}
\textrm{Im}\int_{\zeta_{0}}^{\zeta}(\kappa(\xi)-\kappa(\zeta_{0}))d\xi=0
\end{equation}
These curves are the {\it Stokes lines} beginning at $\zeta_{0}$.
According to equation (\ref{eq:qmdet}), the Stokes line definition
is independent of the choice of the branch of
$\kappa$.\smallpagebreak Assume that $W'(\zeta_{0})\neq 0$. Equation
(\ref{eq:qmsqrt}) implies that there are exactly three Stokes lines
beginning at $\zeta_{0}$. The angle between any two of them at this
point is equal to $\frac{2\pi}{3}$. Indeed for $\zeta$ near
$\zeta_0$, we have
$$
\kappa(\zeta)=\kappa(\zeta_0)+c\sqrt{\zeta-\zeta_0}(1+o(1)).$$So
$$
\int_{\zeta_0}^{\zeta}(\kappa(\zeta)-\kappa(\zeta_0))d\zeta=c(\zeta-\zeta_0)^{3/2}(1+o(1)).$$
\subsubsection{Stokes lines for $E_{0}\in J$}
We describe the Stokes lines beginning at $\varphi_{1}^{-}(E)$ and
$\varphi_{1}^{+}(E)$. Since $W$ is real on $\R$, the set of the
Stokes lines is symmetric with respect to the real line.\newline
First, $\kappa_{1}$ is real on the interval
$[\varphi_{1}^{-}(E),\varphi_{1}^{+}(E)]$; therefore this set is a
Stokes line starting at $\varphi_{1}^{-}(E)$. The two other Stokes
lines beginning at $\varphi_{1}^{-}(E)$ are symmetric with respect
to the real axis. We denote by $\sigma^{-}_{1}(E)$ the Stokes line
going downward and by $\overline{\sigma_{1}}(E)$ its symmetric.
Similarly, we denote by $\sigma_{1}^{+}(E)$ and
$\overline{\sigma_{1}^{+}}(E)$ the two other Stokes lines starting
at $\varphi_{1}^{+}(E)$, $\sigma^{+}_{1}(E)$ goes upward. These
Stokes lines are represented in figure \ref{fig:ls}

\psset{unit=1em,linewidth=.05}

\psset{unit=.7em,linewidth=.1}
\begin{center}
\begin{figure}
\begin{pspicture}(-10,-10)(10,10)
\psline[linewidth=0.005](-10,0)(10,0)
\psline[linewidth=0.05](-10,8)(10,8)
\psline[linewidth=0.05](-10,-8)(10,-8)
\psline(-6.5,-0.2)(-6.5,0.2)\uput[180](-8.5,-1){$\varphi_{1}^{-}(E)$}
\psline(3.5,-0.2)(3.5,0.2)\uput[180](6.5,-1){$\varphi_{1}^{+}(E)$}
\psline(-6.5,0)(3.5,0)
\pscurve(3.5,0)(5.4,5.3)(5.8,8)\uput[180](7.4,4.5){$\sigma_1^{+}(E)$}
\pscurve(3.5,0)(5.4,-5.3)(5.8,-8)\uput[180](7.4,-4.5){$\overline{\sigma_1^{+}(E)}$}
\pscurve(-6.5,0)(-7.5,4)(-7.8,8)\uput[180](-9.5,3.5){$\overline{\sigma_1^{-}(E)}$}
\pscurve(-6.5,0)(-7.5,-4)(-7.8,-8)\uput[180](-9.5,-3.5){$\sigma_1^{-}(E)$}
\end{pspicture}
\caption{Stokes lines for $E\in J$}\label{fig:ls}
\end{figure}
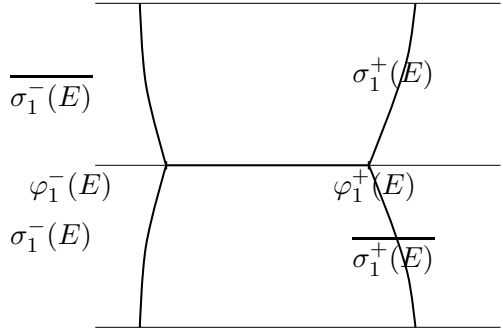
\end{center}
\begin{Le}
\label{le:desc_st_lines} The Stokes lines $\sigma_{1}^{-}(E)$ and
$\sigma_{1}^{+}(E)$ satisfy the following properties:
\begin{itemize}
\item The Stokes lines $\sigma_{1}^{-}(E)$ and $\sigma_{1}^{+}(E)$ stay vertical.
 \item $\sigma_{1}^{-}(E)$ and $\sigma_{1}^{+}(E)$ do
not intersect one another.
\end{itemize}
\end{Le}
The proof of this Lemma is similar to the studies done in
\cite{FEKL1,KM1,Mg,Mg1}. We do not give the details.
\subsection{Construction of a consistent basis near $[\varphi_{1}^{-}(E),\varphi_{1}^{+}(E)]$}
We recall this result, proved in \cite{KM1}.
\begin{Pro}\cite{KM1}\label{pro:consbas} Fix $E_{0}\in
J,$ and $\kappa_1$ a continuous determination of the complex
momentum on $[\varphi_{1}^{-}(E),\varphi_{1}^{+}(E)]$. There exists
a real number $Y>0$, a complex neighborhood $\mathcal{V}_{1}$ of
$E_{0}$ and a consistent basis $(f_{1},(f_{1})^{*})$ of solutions of
equation (\ref{eq:qperbis}) such that $f_{1}$ has the standard
asymptotic behavior:
\begin{equation}
f_{1} \sim
e^{\frac{i}{\varepsilon}\int\tilde{\kappa}_{1}}\Psi_{1}^{+},
\end{equation}
and
$$
f_1^*(x,\zeta,E)=\overline{f(x,\overline{\zeta},\overline{E})}.
$$
to the left of $\sigma_{1}^{-}(E)\cup
\overline{\sigma_{1}^+(\overline{E})}$ (resp. to the right of
$\sigma_{1}^+(E)\cup \overline{\sigma_{1}^-(\overline{E})}$). The
determination $\tilde{\kappa}_{1}$ is the continuation of
$\kappa_{1}$ through $\{\zeta\in (\sigma_{1}^-(E)\cup
\overline{\sigma_{1}^+(\overline{E})})\cap S_{Y}\ ;\
\textrm{Im}\kappa_{1}(\zeta)>0\}$ (resp. through $\{\zeta\in
(\sigma_{1}^-(E)\cup \overline{\sigma_1^+(\overline{E})})\cap S_{Y}\
;\ \textrm{Im} \kappa_{1}(\zeta)<0\}$).
\end{Pro}
We mimic the analysis done in section 5 of \cite{FEKL1}. Precisely,
we start by a local construction of the solution $f$ using canonical
domain; then, we apply continuation tools i.e the rectangle Lemma,
the adjacent domain principle and the Stokes Lemma.
\section{The Proof of Theorem \ref{latouf2}}
The Proof of Theorem \ref{latouf2} follows the same ideas as the
computations  given in section 10.2 of \cite{FEKL1}. Below we only
give the details for the proof of (\ref{eq:moncoeffplus}) and
(\ref{eq:moncoeffmoins}).
\subsection{Strategy of the computation}
\label{sec:technbg} We now begin with the construction of the
consistent basis the monodromy matrix of which we compute. Recall
that $(H1)-(H4)$ are satisfied.\newline In the present section, we
construct and study a solution $f$ of (\ref{eq:dec}) satisfying
\ref{eq:consbis}. \smallpagebreak To use the complex WKB method, we
perform the following change of variable in (\ref{eq:dec})
\begin{equation}\label{eq:newvar}
x-z\rightarrow x,\quad \varepsilon z\rightarrow \zeta.
\end{equation}
Then (\ref{eq:dec}) takes the form (\ref{eq:qperbis}). In the new
variables, the consistency condition (\ref{eq:consbis}) becomes
(\ref{eq:coh}). Note also that in the new variables, for two
solutions to (\ref{eq:qperbis}) to form a consistent basis, in
addition to being a basis of consistent solutions, their Wronskian
has to be independent of $\zeta$.\smallpagebreak Consider the basis
$(f_{1},f_{1}^{*})$ constructed around
$[\varphi_{1}^{-},\varphi_{1}^{+}]$ as in Proposition
\ref{pro:consbas}. Then the monodromy matrix associated to the basis
$(f_1,f_1^{*})$ (defined in section \ref{sec:monmat}) satisfies:
\begin{equation}
\left(\begin{array}cf_{1}(x,\zeta,E,\varepsilon)\\
{f_{1}^{*}(x,\zeta,E,\varepsilon)}\end{array}\right)=M(\zeta,E,\varepsilon)
\left(\begin{array}c{f_{1}(x,\zeta-2\pi,E,\varepsilon)}\\
{f_{1}^{*}(x,\zeta-2\pi,E,\varepsilon)}\end{array}\right).
\end{equation}
The aim of this section is the computation of
$M(\zeta,E,\varepsilon)$. The definition of the monodromy matrix
implies that \begin{equation}
A(\zeta)=M_{1,1}=\frac{w(f_1(x+2\pi,\zeta)f_{1}^*(x,\zeta))}{w(f_1(x,\zeta),f_{1}^*(x,\zeta))},
\
B(\zeta)=M_{12}=\frac{w(f_1(x,\zeta),f_1(x,\zeta+2\pi))}{w(f_1(x,\zeta),f_1^*(x,\zeta))}.\label{dim}\end{equation}
 This gives that the monodromy matrix is analytic in $\zeta$ in the strip
 $S_Y$ and in $E$ in a constant neighborhood of $E_0$. By the definition of $f_1$ we get that $A$ and $B$
 are $\varepsilon$-periodic in $\zeta\in S_Y$. This is a an immediate consequence of the properties of $f_1$.
\newline Therefore, we will compute the Fourier series of $A$ and $B$.The strategy of the computation is based
on the ideas of \cite{FEKL1} and we first recall some notions
presented there. We refer the reader to this paper for more details.
\newline
 Let $h$ and $g$ having a standard asymptotic behavior in regular domains
 $D_h$ and $D_g$ and solutions of (\ref{eq:qperbis}):
 $$
 h\sim e^{\frac{i}{\varepsilon}\int_{\zeta_h}^{\zeta}\kappa_{h}d\zeta}\psi_{h}(x,\zeta),\ {\rm{and}}
 \ g\sim
 e^{\frac{i}{\varepsilon}\int_{\zeta_g}^{\zeta}\kappa_{g}d\zeta}\psi_{g}(x,\zeta).
 $$
 Here, $\kappa_h$, (resp. $\kappa_g$) is an analytic branch of the
 complex momentum in $D_h$ (resp.$D_g$), $\Psi_h$ (resp. $\Psi_g$)
 is the canonical Bloch solution defined on $D_h$ (resp. $D_g$),
 and $\zeta_h$ (resp. $\zeta_g$) is the normalization point for
 $h$ (resp. $g$).\newline
 As the solutions $h$ and $g$ satisfy the consistency condition
 (\ref{eq:coh}), their Wronskian is $\varepsilon$-periodic in
 $\zeta$.

 \subsubsection{Arcs}
  We assume that $D_g\cap D_h$ contains a simply connected
  domain $\tilde{D}$. Let $\gamma$ be a regular curve going from
  $\zeta_g$ to $\zeta_h$ in the following way: staying in $D_g$,
  it goes from $\zeta_g$ to some point in $\tilde{D}$, then,
  staying in $D_h$, it goes to $\zeta_h$. We say that $\gamma $
  is an arc associated to the triple $h,g$ and
  $\tilde{D}$.\newline As $\tilde{D}$ is simply connected, all the
  arcs associated to the triple, $h,\ g$ and $\tilde{D}$. As
  $\tilde{D}$ is simply connected, all the arcs associated to one
  and the same triple can naturally be considered as equivalent; we
  denote them by $\gamma(h,g,\tilde{D})$.
   \newline We continue $\kappa_h$ and $\kappa_g$ analytically
   along $\gamma(h,g,\tilde{D})$. From the properties of
   $\kappa$, we deduce that, for $V$ a small neighborhood of
   $\gamma$, one has
   \begin{equation}
   \kappa_g(\zeta)=\sigma \kappa_h+2\pi m,\ \ \rm{for\ some}\ m\in
   \mathbb{Z},\quad\textrm{and}\quad \sigma\in\{1,-1\}.
   \end{equation}
   $\sigma=\sigma,h,g,\tilde{D}$ is called the signature of
   $\gamma(h,g,\tilde{D})$, and $m=m(h,g,\tilde{D})$ the index of
   $\gamma(h,g,\tilde{D})$.
   \subsubsection{The meeting domain}
   Let $\tilde{D}$ be as above. We call $\tilde{D}$ {\it{a
   meeting domain}}, if, in $\tilde{D}$, the function $\textrm{Im}\ \kappa_h$ and $\textrm{Im}\ \kappa_g$ do not vanish and are of opposite
   signs.\newline Note that, for small values of $\varepsilon$,
   whether $\zeta\mapsto g(x,\zeta)$ and $\zeta\mapsto
   h(x,\zeta)$ increase or decrease is determined by the
   exponential factor $\displaystyle
   e^{\int_{\zeta_g}^{\zeta}}\kappa_gd\zeta$ and $\displaystyle
   e^{\int_{\zeta_h}^{\zeta}}\kappa_hd\zeta$.
   So, roughly, in a meeting domain, along the lines $\textrm{Im}
   \zeta=Const$, the solutions $h$ and $g$ increase in opposite
   directions.
   \subsubsection{The action and the amplitude of an arc}.
   We call the integral
   $$
   S(h,g,\tilde{D})=\int_{\gamma(h,g,\tilde{D})} \kappa_gd\zeta,
   $$
   the {\it{action}} of the arc $\gamma=\gamma(h,g,\widetilde{D})$.
   Clearly, the action takes the same value for equivalent
   arcs.\newline
   Assume that $\mathcal{E}(\zeta)\notin P\cup Q$ along
   $\gamma(h,f,\tilde{D})$. Consider the function
   $q_g=\sqrt{k'(\mathcal{E})}$ and the $1$-form
   $\Omega_g(\mathcal{E}(\zeta))$ in the definition of $\Psi_g$.
   Continue them analytically along $\gamma$. We set
   \begin{equation}
   A(h,g,\gamma)=(\frac{q_g}{q_h})|_{\zeta=\zeta_h}e^{\int_{\gamma}\omega_g}.
   \end{equation}
   $A$ is called the {\it{the amplitude}} of the arc $\gamma$. The
  properties of $\Omega$ imply that the amplitudes of two
  equivalent arcs $\gamma(h,g,\tilde{D})$ coincide.
  \subsection{Results on the Fourier coefficients}
We recall the following result from \cite{FEKL1}
\begin{Pro}\label{oua2}
Let $d=d(h,g)$ be a meeting domain for $h$ and $g$, and $m=m(h,g,d)$
be the corresponding index. Then

\begin{equation}
w(h,g)=w_me^{\frac{2\pi im}{\varepsilon}(\zeta-\zeta_h)}(1+o(1));\
\zeta\in S(d),\label{ha}
\end{equation}
where $w_m$ is the constant given by
\begin{equation}
w_m=A(h,g,d)e^{\frac{i}{\varepsilon}S(h,g,d)}w(\psi_{+}(\cdot,\zeta_h),\psi_{-}(\cdot,\zeta_h)),
\end{equation}
Here $\psi_+=\psi_h$ and $\psi_{-}$ is complementary to $\psi_+$.
The asymptotic (\ref{ha}) is uniform in $\zeta$ and $E$ when $\zeta$
stays in a fixed compact of $S(d)$ and $E$ in a small enough
neighborhood of $E_0$.
\end{Pro}

\subsubsection{The index $m$ }
Let $\zeta_0$ be a regular point. consider a regular curve $\gamma$
going from $\zeta_0$ to $\zeta_0+2\pi$. Let $\kappa$ be a branch of
the complex momentum that is continuous on $\gamma$. We call the
couple $(\gamma,\kappa)$ a period. Let $(\gamma_1,\kappa_1)$ and
$(\gamma_2,\kappa_2)$ be two periods. Assume that one can
continuously deform $\gamma_1$ into $\gamma_2$ without intersecting
any branching point. By this we define an analytic continuation of
$\kappa_1$ to $\gamma_2$.
If the analytic continuation coincide with $\kappa_2$, we say that the periods are equivalent.\\
Consider the branch $\kappa$ along the curve $\gamma$ of a period
$(\gamma,\kappa)$. In a neighborhood of $\zeta_0$, the starting
point $\gamma$, one has
$$
\kappa(\zeta+2\pi)=\sigma\kappa(\zeta)+2\pi m,\quad\quad
\sigma\in\{\mp 1\},\quad\quad m\in \mathbb{Z}.
$$
The numbers $\sigma=\sigma(\gamma,\kappa)$ and $m=m(\gamma,\kappa)$
are called respectively the signature and
 the index of the period $(\gamma,\kappa)$. They coincide for equivalent periods.

Recall that
$$
\mathcal{G}=\cup_{k\in \mathbb{Z}}\{\cup _{j=1}^N\mathcal{G}_j+2\pi
k\emph{}\};
$$
is the pre-image with respect to $\mathcal{E}$ of the union of the
spectral gaps of $H_0$. One has
\begin{Le}\cite{FEKL2}\label{sar2}
Let $(\gamma,\kappa)$ be a period such that $\gamma$ starts at a
point $\zeta_0\notin \mathcal{G}$. Assume that $\gamma$ intersects
$\mathcal{G}$ exactly $n$ times ($n\in \mathbb{N}$) and that at all
intersection points, $W'\neq 0$. Let $r_1,\cdots, r_n$ be the values
that $\textrm{Re}\ \kappa$ takes consecutively at these intersection
points as $\zeta$ moves along $\gamma$ from $\zeta_0$ to
$\zeta_0+2\pi$. Then,
\begin{eqnarray}
\sigma(\gamma,\kappa)=(-1)^n,\
m(\gamma,\kappa)&=&\frac{1}{\pi}(r_n-r_{n-1}+r_{n-2}-\cdots+(-1)^{n-1}r_1),\\
&=&\frac{(-1)^{n-1}}{\pi}(r_1-r_2+\cdots +(-1)^{n-1}r_n).
\end{eqnarray}
\end{Le}
\subsection{The Fourier coefficients}
\subsubsection{For $B$}
By equation (\ref{dim}), we have to compute $w
(f(\cdot,\zeta),f(\cdot,\zeta+2\pi))$. With this aim in view, we
apply the construction done in section \ref{sec:technbg} with:
$$h(x,\zeta)=f(x,\zeta),\ g(x,\zeta)=(Tf)(x,\zeta),\ \ {\rm{with}}\ \ \ (Tf)(x,\zeta)=f(x,\zeta+2\pi);$$
and
\begin{equation}D_h=\mathcal{D},D_g=\mathcal{D}-2\pi,\ \zeta_h=\zeta_0,\ \ \zeta_g=\zeta_0-2\pi,\label{oua}\end{equation}
\begin{equation}
\forall \zeta\in D_h,\ \kappa_h(\zeta)=\kappa(\zeta),\ \
\forall\zeta\in D_g,\ \
\kappa_g(\zeta)=\kappa(\zeta+2\pi).\label{sar1}
\end{equation}We will start by
\subsubsection{Above the real line}
We take the meeting domain $D_0$ as the subdomain of the strip
$0<\textrm{Im} \zeta<Y$ between the Stokes lines $\sigma_1^+-2\pi$
and $\sigma_1^+$.(In this domain we have $\textrm{Im}\
\kappa_g=-\textrm{Im}\ \kappa_h<0$). The arc $\gamma_0$ connects the
point $\zeta_g$ to $\zeta_h$. By (\ref{oua}), this defines the
period $(\gamma_0,\kappa_g)$. Using (\ref{sar1}), one gets that
$m(f(\cdot,\zeta),f(\cdot,\zeta+2\pi, D_0))=m(\gamma_0+2\pi,\kappa)$
\cite{FEKL1}.\newline We use Lemma \ref{sar2} to compute the index.
To do this, we have to compute $\textrm{Re}\ \kappa$ at the
intersection of $\gamma_0+2\pi$ and $\mathcal{G}$.\newline As $\zeta
\to \textrm{Re}\ \kappa(\zeta)$ is constant on any connected
component $\mathcal{G}_j$ of $\mathcal{G}$. Let us start by defining
the index $\lambda_j^{+}$ of $\mathcal{G}_j$, the result of
alternation $(\cdots ,+,-,+,-,\cdots)$ due to
 the crossing of $\mathcal{G}_j$. We notice that $\lambda_{j}^{+}\in \{-1,0,1\}$. \newline
 We set
 \begin{equation}N_{j}=1+\sum_{i=1}^{j-1}(\sum_{l=1}^{n_{i}}(o_l-1)).\label{recto}\end{equation}
Here $n_j$ is the number of extremum in $\mathcal{G}_j$, and $o_l$
is the order of the $l^{\text{th}}$ extremum. The following
relations hold
\begin{equation}
\lambda_1^{+}=\frac{1+(-1)^{\sum_{l=1}^{n_1}(o_l-1)}}{2}
\end{equation} and for $ 2\leq j\leq N$,
\begin{equation}\lambda_j^{+}=(-1)^{N_j}\Big(\frac{-1+(-1)^{\sum_{l=1}^{n_j}(o_l-1)}}{2}\Big).
\end{equation}
Without loss of generality we assume that $\textrm{Re}\ \kappa=0$ on
$(0,\varphi^-_1)$. By the above notation, we get that
\begin{equation}
m(\gamma_0,\kappa)=(-1)^{n}\Big(p_1\lambda_1^{+}+(p_1+p_2)\lambda_2^{+}+\cdots+(\sum_{i=1}^jp_i)\lambda_j^{+}+\cdots
+(\sum_{i=1}^{N}p_i)\lambda_{N}^{+}\Big)=P_+.
\end{equation}
Here $p_i$ is the index of $\mathcal{B}_i$ and $n=N_N$.\newline Now
using (\ref{dim}) for $B$ and Proposition \ref{oua2} we get
\begin{equation}
B=A(f,T(f),\tilde{D}_0)e^{\frac{i}{\varepsilon}S(f,T(f),\tilde{D}_0)}
\cdot e^{\frac{-2i\pi
P_+\zeta_0}{\varepsilon}}=B_{+}(E,\varepsilon)\cdot
e^{-\frac{i}{2\pi\varepsilon}P_+\zeta_0}, \
T(f)(x,\zeta)=f(x,\zeta+2\pi).
\end{equation}
\subsubsection{Below the real line}

\psset{unit=1em,linewidth=.05}

\psset{unit=.5em,linewidth=.1}
\begin{center}
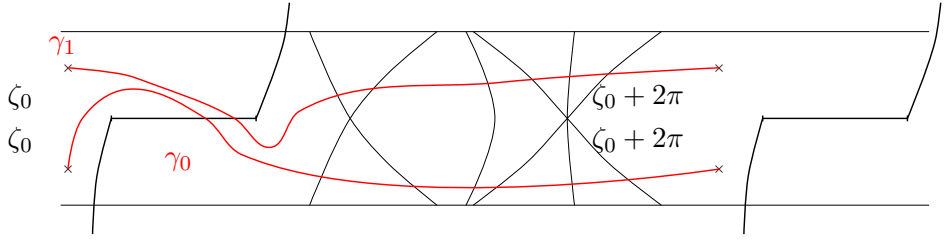
\begin{figure}
\begin{pspicture}(-20,-10)(60,10)
\psline[linewidth=0.01](-10,6)(50,6)
\psline[linewidth=0.01](-10,-6)(50,-6)
\pscurve[linewidth=0.01](18.5,-6)(22.25,-3)(25,0)(27.75,3)(31.5,6)\pscurve[linewidth=0.01](25.5,-6)(25,0)(25.5,6)\pscurve[linewidth=0.01](31.5,-6)(27.75,-3)(25,0)(22.25,3)(18.5,6)
\pscurve[linewidth=0.01](18,-6)(20,0)(18,6)
\pscurve[linewidth=0.01](16,-6)(12.5,-3)(10,0)(7.2,6)\pscurve[linewidth=0.01](7.2,-6)(10,0)(12.5,3)(16,6)
\psdots[dotstyle=x](-9.5,3.5)\uput[180](-11,1.5){$\zeta_{0}$}
\psline(-6.5,-0.2)(-6.5,0.2)
\psline(3.5,-0.2)(3.5,0.2)
\psline(-6.5,0)(3.5,0)
\pscurve(3.5,0)(5.4,5.3)(5.8,8)
\pscurve(-6.5,0)(-7.5,-4)(-7.8,-8)
\psline(38.5,-0.2)(38.5,0.2) \psline(48.5,-0.2)(48.5,0.2)
\psline(38.5,0)(48.5,0) \pscurve(48.5,0)(50.4,5.3)(50.8,8)
\psdots[dotstyle=x](35.5,3.5)\uput[180](34,1.5){$\zeta_{0}+2\pi$}
\psdots[dotstyle=x](35.5,-3.5)\uput[180](34,-1.5){$\zeta_{0}+2\pi$}
\psdots[dotstyle=x](-9.5,-3.5)\uput[180](-11,-1.5){$\zeta_{0}$}
\pscurve(38.5,0)(37.5,-4)(37.2,-8)
\pscurve[linecolor=red](-9.5,3.5)(-5.5,3)(-4,2.5)(2,0)(4.5,-2)(6.5,0.5)(20.5,2.5)(35.5,3.5)
\uput[180](-8,5){\textcolor{red}{$\gamma_1$}}
\pscurve[linecolor=red](-9.5,-3.5)(-8.5,0)(-5.5,2)(0,0)(2.5,-2.5)(35.5,-3.5)
\uput[180](0,-3){\textcolor{red}{$\gamma_0$}}

\end{pspicture}
\caption{Periods equivalent to $\gamma_0$ and
$\gamma_1$}\label{fig:gamma0}
\end{figure}
\end{center}

Below the real line, we take the domain of the strip $-Y<
\textrm{Im} \zeta<0$ located between the stokes line
$\sigma_1^--2\pi $ and $\sigma_1^-$ as a regular domain; which we
denote by $\tilde{D_0}$. We set
$\tilde{\gamma}_0=\gamma(f,T(f),\tilde{D}_1)$, then it defines a
period, and so
$m(f,T(f),\tilde{D}_1)=m(\tilde{\gamma}_0+2\pi,\kappa)$. The curve
defining a period equivalent to $(\tilde{\gamma}_0+2\pi,\kappa)$ is
represented in figure \ref{fig:gamma0}.\newline  The computation of
the index $\lambda^-_{j}$ of $\mathcal{G}_j$ gives in this case that
$$\lambda^{-}_{1}=\frac{-1+(-1)^{\sum_{i=1}^{n_1}(o_l-1)}}{2},$$
and
$$
\lambda^{-}_{j}=(-1)^{N_j}\Big(\frac{-1+(-1)^{\sum_{l=1}^{l=n_j}(o_l-1)}}{2}\Big).
$$
By this notation we get that
\begin{equation}
m(\gamma_0,\kappa)=(-1)^{n}\Big(p_1\lambda_1^{-}+(p_1+p_2)\lambda_2^{-}+\cdots+(\sum_{i=1}^jp_i)\lambda_j^{-}+\cdots
+(\sum_{i=1}^{N}p_i)\lambda_{N}^{-}\Big)=P_-.
\end{equation}
Using (\ref{dim}) for $B$ and Proposition \ref{oua2} we get
\begin{equation}
B_{P_-}=A(f,T(f),\tilde{D}_1)e^{\frac{i}{\varepsilon}(S(f,Tf,\tilde{D}_1)-2\pi
P_-\zeta_0)}=B_-(E,\varepsilon)\cdot e^{\frac{i}{\varepsilon}2\pi
P_-\zeta_0},\ \ (Tf)(x,\zeta)=f(x,\zeta+2\pi).
\end{equation}
\subsection{For $A$}
For the computation of $A$ using equation (\ref{dim}), we have to
compute $w (f^*(\cdot,\zeta),f(\cdot,\zeta+2\pi))$. It suffices to
apply the method presented in section \ref{sec:technbg} with:
$$h(x,\zeta)=f^*(x,\zeta),\ g(x,\zeta)=(Tf)(x,\zeta),\ \ {\rm{with}}\ \ \ (Tf)(x,\zeta)=f(x,\zeta+2\pi);$$
and
\begin{equation}D_h=\mathcal{D}^*,D_g=\mathcal{D}-2\pi,\ \zeta_h=\zeta_0,\ \ \zeta_g=\zeta_0-2\pi,\label{yous}\end{equation}
\begin{equation}
\forall \zeta\in D_h,\
\kappa_h(\zeta)=-\overline{\kappa}(\overline{\zeta}),\ \
\forall\zeta\in D_g,\ \
\kappa_g(\zeta)=\kappa(\zeta+2\pi).\label{hal}
\end{equation}
\subsubsection{Above the real line}

In this case, $\tilde{D}_0$, the meeting domain, is the subdomain of
the strip $\displaystyle \{0<\textrm{Im} \zeta <Y\}$ located between
the lines $\sigma_1^+-2\pi$ and $\overline{\sigma_1^-}$, the
symmetric to $\sigma_1^-$ with respect to $\mathbb{R}$. The arc
$\gamma(f^*,T(f),\tilde{D}_0) $ defines a period
$(\tilde{\gamma}_0,\kappa_g)$, in figure \ref{fig:gammabis} we
represent the curve $\tilde{\gamma}_0+2\pi$.\newline Similarly to
the computation of $B$, we define the index $\beta^+_{j}$ of
$\mathcal{G}_j$ and in this case, we have:
$$\beta^{+}_{1}=\frac{-1+(-1)^{\sum_{i=1}^{n_1}(o_l-1)}}{2},$$
and
$$
\beta^{+}_{j}=(-1)^{N_j}\Big(\frac{-1+(-1)^{\sum_{l=1}^{l=n_j}(o_l-1)}}{2}\Big).
$$
By this notation we get that
\begin{equation}
m(\gamma_0,\kappa)=(-1)^{n}\Big(p_1\beta_1^{+}+(p_1+p_2)\beta_2^{+}+\cdots+(\sum_{i=1}^jp_i)\beta_j^{+}+\cdots
+(\sum_{i=1}^{N}p_i)\beta_{N}^{+}\Big)=Q_+.
\end{equation}
Using (\ref{dim}) for $A$ and Proposition \ref{oua2} we get that
$$m(f^*,Tf,\tilde{D}_0)=m(\tilde{\gamma}_0,\kappa)=Q_+,$$
and
\begin{equation}A=A(f^*,Tf,\tilde{D}_0)e^{\frac{i}{\varepsilon}(S(f^*,Tf,\tilde{D}_0)-2\pi
Q_+\zeta_0)}=A_+(E,\varepsilon)\cdot e^{-\frac{i}{\varepsilon}2\pi
Q_+},\ \ \ (Tf)(x,\zeta)=f(x,\zeta+2\pi).
\end{equation}
\psset{unit=1em,linewidth=.05}

\psset{unit=.5em,linewidth=.1}
\begin{center}
\begin{figure}
\begin{pspicture}(-20,-10)(60,10)
\psline[linewidth=0.01](-10,6)(50,6)
\psline[linewidth=0.01](-10,-6)(50,-6)
\pscurve[linewidth=0.01](18.5,-6)(22.25,-3)(25,0)(27.75,3)(31.5,6)\pscurve[linewidth=0.01](25.5,-6)(25,0)(25.5,6)\pscurve[linewidth=0.01](31.5,-6)(27.75,-3)(25,0)(22.25,3)(18.5,6)
\pscurve[linewidth=0.01](18,-6)(20,0)(18,6)
\pscurve[linewidth=0.01](16,-6)(12.5,-3)(10,0)(7.2,6)\pscurve[linewidth=0.01](7.2,-6)(10,0)(12.5,3)(16,6)
\psdots[dotstyle=x](-9.5,-5)\uput[180](-11,-3){$\zeta_{0}$}
\psline(-6.5,-0.2)(-6.5,0.2)
\psline(3.5,-0.2)(3.5,0.2)
\psline(-6.5,0)(3.5,0)
\pscurve(3.5,0)(5.4,-5.3)(5.8,-8)
\pscurve(-6.5,0)(-7.5,4)(-7.8,8)
\psline(38.5,-0.2)(38.5,0.2) \psline(48.5,-0.2)(48.5,0.2)
\psline(38.5,0)(48.5,0) \pscurve(48.5,0)(50.4,5.3)(50.8,8)
\psdots[dotstyle=x](-9.5,3.5)\uput[180](-11,1.5){$\zeta_{0}$}
\psdots[dotstyle=x](35.5,-5)\uput[180](34,-3){$\zeta_{0}+2\pi$}
\psdots[dotstyle=x](35.5,3.5)\uput[180](34,1.5){$\zeta_{0}+2\pi$}
\pscurve(38.5,0)(37.5,-4)(37.2,-8)
\pscurve[linecolor=red](-9.5,3.5)(-5.5,-2)(-4,0)(-2.5,2)(20.5,2.5)(35.5,3.5)
\uput[180](0,5){\textcolor{red}{$\widetilde{\gamma}_1$}}
\pscurve[linecolor=red](-9.5,-5)(0,0)(2.5,2.5)(7,-2)(35.5,-5)
\uput[180](0,-3){\textcolor{red}{$\widetilde{\gamma}_0$}}
\end{pspicture}
\caption{Periods equivalent to $\widetilde{\gamma}_0$ and
$\widetilde{\gamma}_1$ }\label{fig:gammabis}
\end{figure}
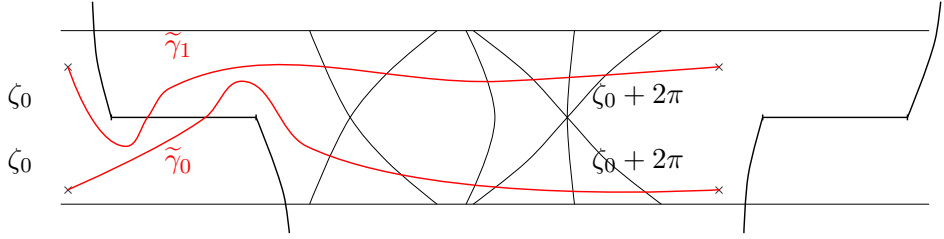
\end{center}
\subsubsection{Below the real line}
In this case, $\tilde{D}_1$, the meeting domain, is the subdomain of
the band $\displaystyle \{ -Y<\textrm{Im} \zeta <0\}$ located
 between the lines $\overline{\sigma_1^+}$, symmetric of $\sigma_1^+$ with
 respect to $\mathbb{R}$, and $\sigma_1^--2\pi$. The arc $\gamma(h,g,\tilde{D}_1)$
  defines a period $(\tilde{\gamma}_1,\kappa_g)$; the curve $\tilde{\gamma}_1+2\pi$
   is represented in figure \ref{fig:gammabis}.
  One obtains that
  \begin{equation}
\beta_1^{-}=\frac{1+(-1)^{\sum_{l=1}^{n_1}(o_l-1)}}{2},
\end{equation}and for $ 2\leq j\leq N$,
\begin{equation}\beta_j^{-}=(-1)^{N_j}\Big(\frac{-1+(-1)^{\sum_{l=1}^{n_j}(o_l-1)}}{2}\Big).
\end{equation}
 So by the above notation we get that
\begin{equation}
m(\gamma_0,\kappa)=(-1)^{n}\Big(p_1\beta_1^{-}+(p_1+p_2)\beta_2^{-}+\cdots+(\sum_{i=1}^jp_i)\beta_j^{-}+\cdots
+(\sum_{i=1}^{N}p_i)\beta_{N}^{-}\Big)=Q_-
\end{equation}
  One obtains that $m(f^*,T(f),\tilde{D}_1)=m(\tilde{\gamma}_1+2\pi,\kappa)=Q_-$,
  and
  \begin{equation}A=A(f^*,Tf,\tilde{D}_1)e^{\frac{i}{\varepsilon}(S(f^*,Tf,\tilde{D}_1)-2\pi
Q_-\zeta_0)}=A_-(E,\varepsilon)\cdot e^{\frac{i}{\varepsilon}2\pi
Q_-\zeta_0}\ ;\ \ (Tf)(x,\zeta)=f(x,\zeta+2\pi).
\end{equation}
\newline \textit{$\mathbf{Acknowledgements.}$ The authors would like to
thank F. Klopp for proposing this problem and many valuable comments
and remarks. H.N thanks M. Dimassi for suggesting some interesting
references}
\bibliographystyle{plain}

\end{document}